\documentclass[twocolumn,aps,pra,amsmath,amssymb,superscriptaddress,bibnotes]{revtex4-1}

\usepackage{graphicx}
\usepackage{dcolumn}
\usepackage{bm}
\usepackage{hyperref,soul}

\usepackage{epstopdf}
\usepackage{times,txfonts}

\usepackage{color,bbm}
\newcommand{\beq}{\begin{equation}}
\newcommand{\eeq}{\end{equation}}
\newcommand{\bqa}{\begin{eqnarray}}
\newcommand{\eqa}{\end{eqnarray}}

\definecolor{maroon}{rgb}{0.7,0,0}

\definecolor{ngreen}{rgb}{0.3,0.7,0.3}

\definecolor{golden}{rgb}{0.8,0.6,0.1}

\begin{document}

\title{Demonstrating quantum coherence and metrology that is resilient to transversal noise}

\author{Chao Zhang}
\affiliation{Key Laboratory of Quantum Information, University of Science and Technology of China, CAS, Hefei, 230026, People's Republic of China}
\affiliation{CAS Center For Excellence in Quantum Information and Quantum Physics, University of Science and Technology of China, Hefei, 230026, People's Republic of China}

\author{Thomas R. Bromley}
\email{thomas.r.bromley@gmail.com}
\affiliation{School of Mathematical Sciences and Centre for the Mathematics and Theoretical Physics of Quantum Non-Equilibrium Systems, University of Nottingham, University Park Campus, Nottingham NG7 2RD, United Kingdom}

\author{Yun-Feng Huang}
\email{hyf@ustc.edu.cn}
\affiliation{Key Laboratory of Quantum Information, University of Science and Technology of China, CAS, Hefei, 230026, People's Republic of China}
\affiliation{CAS Center For Excellence in Quantum Information and Quantum Physics, University of Science and Technology of China, Hefei, 230026, People's Republic of China}

\author{Huan Cao}
\affiliation{Key Laboratory of Quantum Information, University of Science and Technology of China, CAS, Hefei, 230026, People's Republic of China}
\affiliation{CAS Center For Excellence in Quantum Information and Quantum Physics, University of Science and Technology of China, Hefei, 230026, People's Republic of China}

\author{Wei-Min Lv}
\affiliation{Key Laboratory of Quantum Information, University of Science and Technology of China, CAS, Hefei, 230026, People's Republic of China}
\affiliation{CAS Center For Excellence in Quantum Information and Quantum Physics, University of Science and Technology of China, Hefei, 230026, People's Republic of China}

\author{Bi-Heng Liu}
\affiliation{Key Laboratory of Quantum Information, University of Science and Technology of China, CAS, Hefei, 230026, People's Republic of China}
\affiliation{CAS Center For Excellence in Quantum Information and Quantum Physics, University of Science and Technology of China, Hefei, 230026, People's Republic of China}

\author{Chuan-Feng Li}
\email{cfli@ustc.edu.cn}
\affiliation{Key Laboratory of Quantum Information, University of Science and Technology of China, CAS, Hefei, 230026, People's Republic of China}
\affiliation{CAS Center For Excellence in Quantum Information and Quantum Physics, University of Science and Technology of China, Hefei, 230026, People's Republic of China}

\author{Guang-Can Guo}
\affiliation{Key Laboratory of Quantum Information, University of Science and Technology of China, CAS, Hefei, 230026, People's Republic of China}
\affiliation{CAS Center For Excellence in Quantum Information and Quantum Physics, University of Science and Technology of China, Hefei, 230026, People's Republic of China}

\author{Marco Cianciaruso}
\email{cianciaruso.marco@gmail.com}
\affiliation{School of Mathematical Sciences and Centre for the Mathematics and Theoretical Physics of Quantum Non-Equilibrium Systems, University of Nottingham, University Park Campus, Nottingham NG7 2RD, United Kingdom}

\author{Gerardo Adesso}
\email{gerardo.adesso@nottingham.ac.uk}
\affiliation{School of Mathematical Sciences and Centre for the Mathematics and Theoretical Physics of Quantum Non-Equilibrium Systems, University of Nottingham, University Park Campus, Nottingham NG7 2RD, United Kingdom}

\date{\today}

\begin{abstract}
Quantum systems can be exploited for disruptive technologies but in practice quantum features are fragile due to noisy environments. Quantum coherence, a fundamental such feature, is a basis-dependent property that is known to exhibit a resilience to certain types of Markovian noise. Yet, it is still unclear whether this resilience can be relevant in practical tasks. Here, we experimentally investigate the resilient effect of quantum coherence in a photonic Greenberger-Horne-Zeilinger state under Markovian bit-flip noise, and explore its applications in a noisy metrology scenario. In particular, using up to six-qubit probes, we demonstrate that the standard quantum limit can be outperformed under a transversal noise strength of approximately equal magnitude to the signal, providing experimental evidence of metrological advantage even in the presence of uncorrelated Markovian noise. This work highlights the important role of passive control in noisy quantum hardware, which can act as a low-overhead complement to more traditional approaches such as quantum error correction, thus impacting on the deployment of quantum technologies in real-world settings.
\end{abstract}


\maketitle

{\it Introduction.---}Harnessing quantum effects holds the promise of revolutionizing information processing in ways that greatly surpass current approaches, including quantum computing, communication, and metrology~\cite{QR}. However quantum resources are very fragile and practical realizations of quantum sensors and processors inevitably interact with their surroundings, eventually losing their nonclassical properties. In particular, the process of ``decoherence''~\cite{Decoherence} stands as one of the major obstacles in realizing scalable quantum technologies. During the past two decades, numerous efforts have been invested to devise active noise control schemes~\cite{QEC1,QEC2,QControl,DD}. Quantum error correction with feedback control~\cite{QEC1,QEC2} provides the most promising scheme to combat arbitrary noise, however the excessive resource overhead keeps it beyond reach of current technology. A complementary approach is to develop passive noise control schemes, which are more affordable, by harnessing the natural resilience of quantum resources to specific noise. For example, placing a system in a decoherence-free subspace (DFS) can make it inherently immune to collective noises~\cite{DFS0,DFS1,DFS2}.

Quantum coherence, encapsulating the idea of superposition of quantum states, is a defining feature of quantum mechanics and also a crucial resource for quantum information processing~\cite{Cohres}. Recently, the development of a rigorous resource framework for coherence~\cite{frame,frame2,frame3} has brought it back to the limelight and motivated a number of studies~\cite{Study1,Study2,Study3,Study4,freezing1}. Coherence is defined with respect to a particular reference basis, usually specified by the physics of the system under investigation~\cite{frame}. As such, one may intuitively expect its resilience to depend on the direction along which the noise acts. Surprisingly, it has been observed that, under suitable conditions, the coherence  in a multi-qubit system (with respect to the computational basis) can remain exactly \emph{constant} under independent bit-flip noise acting on each qubit~\cite{freezing1,freezing2,freezing3,freezingNMR}, in a process known as ``freezing''. This freezing phenomenon takes place despite the quantum state itself evolving due to the noise, highlighting a key difference to the DFS scenario.

It is intriguing to explore practical applications of frozen or more generally {\em resilient} coherence, particularly in quantum parameter estimation~\cite{metrology,metrologyrev1,metrologyrev2}, for which coherence in the eigenbasis of the parameter-imprinting generator is an essential resource. It is well known that the precision of noise-free quantum metrology can beat the standard quantum limit (SQL) and achieve the Heisenberg limit (HL) by exploiting entangled probes, e.g., Greenberger-Horne-Zeilinger (GHZ) states. However, the quantum advantage is much more elusive in realistic environments in which the noise and the unitary evolution imprinting the parameter act on the probes simultaneously. In fact, there are a number of no-go results demonstrating that for most types of uncorrelated noise the asymptotic scaling is constrained to be SQL-like~\cite{parallel,nogo1,nogo2,nogo3,nogo4}. Nevertheless, it has been shown theoretically that, when the noise is concentrated along a direction perpendicular to the unitary dynamics (known as transversal noise), even if the noise is purely Markovian, a superclassical precision scaling in frequency estimation can be maintained by optimizing the interrogation-time~\cite{noisymetrology1,noisymetrology2}. Note that for parallel Markovian noise, the uncorrelated and GHZ probes achieve exactly the same precision, thus no quantum advantage can be achieved~\cite{parallel}.

In this Letter, we use a highly controllable photonic system as an experimental testbed to investigate the resilience of quantum coherence and metrology against transversal noise.
We first demonstrate frozen quantum coherence in a 4-photon GHZ state prepared in both the computational and $\sigma_x$ bases and then subjected to Markovian bit-flip noise. We observe that the quantum Fisher information for estimating a phase encoded along the $\sigma_{z}$ basis is also frozen in the GHZ state prepared in the $\sigma_x$ basis. We then consider a frequency estimation task with additional bit-flip noise, which mimics a scenario of relevance for atomic magnetometry~\cite{magnetometry}.
We demonstrate that the SQL can be surpassed using GHZ probes of up to 6 qubits, despite their exposure to noise of comparable strength to the signal.

\begin{figure}[tb]
\centering
\includegraphics[width=0.5\textwidth]{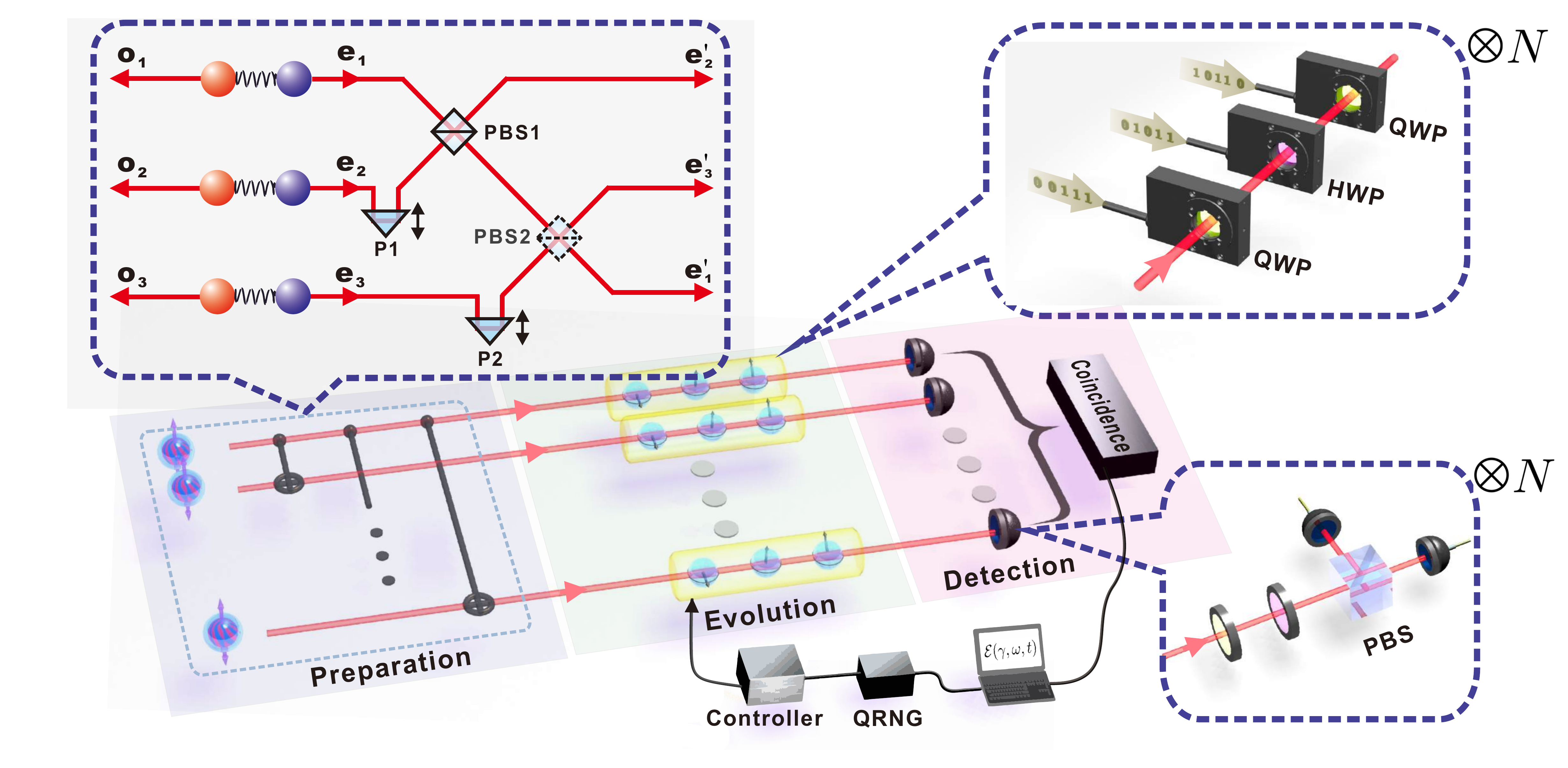}
\vspace*{-.6cm}
\caption{\label{Fig:1} The experimental setup consists of three main steps: preparation, evolution and detection. Ultraviolet laser pulses with a central wavelength of 390nm, pulse duration of 140fs, and a repetition rate of 76 MHz pass through three beamlike type-II spontaneous parametric down-conversion sources (not shown in the figure), and generate three EPR photon pairs in the state of $(|HH\rangle+|VV\rangle)/\sqrt{2}$.
The three e-photons are fed into a multiphoton interferometer consisting of two PBSs. With post-selection, a 6-photon GHZ state can be projected. By removing PBS2, the first two sources generate a 4-photon GHZ state, and the 3-photon GHZ state can be generated by projecting one photon onto $\frac{1}{\sqrt{2}}(|H\rangle+|V\rangle)$ (See SM.IB~\cite{SM} for detail). After preparation, each photon passes through a channel which consists of three randomly rotated wave plates. All the wave plates are mounted on motorized rotation stages and controlled by a quantum random number generator (QRNG). By setting the parameters controlling the wave plates, various sources of noise can be simulated.
In the measurement part, each photon is passed through a narrowband filter and detected by a polarization analysis system. \vspace*{-.3cm}}  \end{figure}

{\it Frozen quantum coherence and quantum Fisher information (QFI).---}Coherence is marked by the presence of off-diagonal elements of a density matrix with respect to a particular basis. Incoherent states are classical mixtures with respect to the basis, corresponding to the set of diagonal density matrices. Given the set of incoherent states $\mathcal{I}$, the degree of coherence of a state $\rho$ can be quantified by how distinguishable $\rho$ is from $\mathcal{I}$, where a distance-based measure can be used to quantify distinguishability (see the Supplemental Material (SM.IIA)~\cite{SM} for further details).

It is important to study the dynamical evolution of coherence quantifiers given the inevitable interaction of quantum systems with their environments. Refs.~\cite{freezing1,freezing2} identified dynamical conditions under which all distance-based coherence monotones can be frozen in a class of $N$-qubit states with maximally mixed marginals ($M_N^3$ states). Time-invariant coherence has been demonstrated under these conditions in a NMR experiment~\cite{freezingNMR}. Subsequently, it has been found that the relative entropy measure of coherence plays a special role in determining freezing conditions, since all coherence monotones are frozen if and only if the relative entropy  is frozen~\cite{freezing3}.
Such a criterion can help us to identify other classes of initial states exhibiting frozen coherence.

GHZ states are widely used as resources for quantum information processing. This paper investigates the dynamical conditions and applications of frozen coherence in $N$-qubit GHZ states, forming a complementary set to the canonical $M_N^3$ states for $N> 2$. We first focus on $4$-qubit GHZ state, prepared in either the computational $\sigma_z$ ($0/1$) basis $|G_4\rangle=\frac{1}{\sqrt{2}}(|0\rangle^{\otimes4}+|1\rangle^{\otimes4})$ or $\sigma_x$ ($\pm$) basis $|G_4^{\pm}\rangle=\frac{1}{\sqrt{2}}(|+\rangle^{\otimes4}+|-\rangle^{\otimes4})$, where $|\pm\rangle=(|0\rangle\pm |1\rangle)/\sqrt{2}$. We see in the following that these states can exhibit both frozen coherence and frozen QFI.

The GHZ states (preparation part in Fig.~\ref{Fig:1}) are generated by combining two sandwich-like EPR photon pairs~\cite{source} through a polarizing beamsplitter (PBS)~\cite{foursource}. Both photon pairs are prepared in the state $\frac{1}{\sqrt{2}}(|HH\rangle+|VV\rangle)$, where H (V) denotes the horizontal (vertical) polarization of photons and encodes the qubit values 0 (1). The fidelity of the prepared GHZ state is as high as $97.5\%$ (see SM.IC~\cite{SM} for further details). Hadamard gates, implemented as half-wave-plates (HWPs) at 22.5$^{\circ}$, can be used to transform the state $|G_4\rangle$ into $|G_4^{\pm}\rangle$. Then, each photon is fed into a bit-flip noise channel (evolution part in Fig.~\ref{Fig:1}, see SM.IIB~\cite{SM} for details).

\begin{figure*}[tb]
\centering
\includegraphics[width=0.95\textwidth]{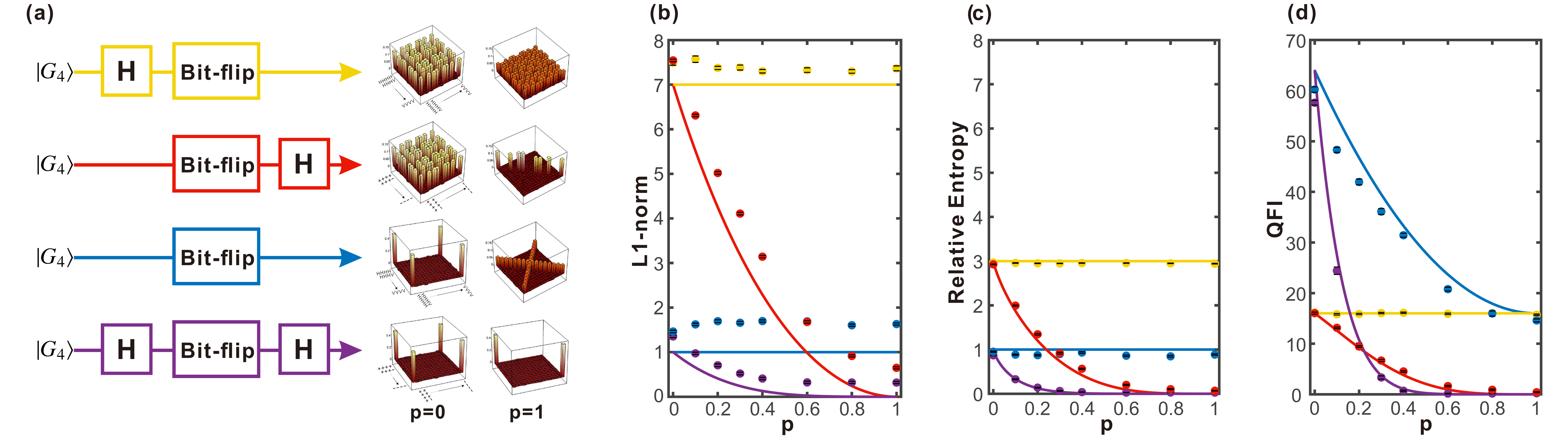}
\vspace*{-.3cm}
\caption{\label{Fig:2} (a) The schematic diagram of the four cases studied in our experiment: the yellow color (first line) for the case that the GHZ state is prepared in the $\pm$ basis and coherence is calculated in the 0/1 basis, denoted $\{\pm, 0/1\}$; red (second line) for $\{0/1, \pm\}$; blue (third line) for $\{0/1, 0/1\}$ and purple (fourth line) for $\{\pm, \pm\}$. We use a Hadamard gate (H) to represent the change between the 0/1 and $\pm$ bases. Shown on the right of the diagram are the real part of the reconstructed density matrices in the corresponding basis for $p=0$ and $p=1$, respectively. The right three panels show the measurement results (dots) of: (b) the $l_1$-norm of coherence, (c) the relative entropy of coherence, and (d) the QFI, along with theoretical predictions (solid lines) for the GHZ states during the evolution. The error bars are smaller than the marker size (the measured $l_1$-norm coherence is always larger than the theoretical value, due to the accumulation of the statistical errors for all off-diagonal elements of the density matrix).
\vspace*{-.3cm}
} \end{figure*}

At the output (detection part in Fig.~\ref{Fig:1}), we perform full state tomography of the evolved states with different noise strengths, allowing a full analysis of the evolution. For each state, we calculate the $l_1$ norm of coherence and the relative entropy of coherence (SM.IIA~\cite{SM}) with respect to both 0/1 basis and $\pm$ basis, resulting in four quantities overall. Figure~\ref{Fig:2}(a) illustrates each case, while Figs.\ref{Fig:2}(b) and (c) show the corresponding coherence dynamics.

When coherence is measured in the computational basis (yellow and blue lines), both coherence quantifiers remain constant for any value of noise strength $p$. On the other hand, the coherence measures in the $\pm$ basis (red and purple lines) decay monotonically to zero. Note that the computational basis forms the eigenbasis of $\sigma_{z}$, which is orthogonal to the bit-flip noise generated by $\sigma_{x}$ (i.e., $\mbox{Tr}(\sigma_{x}\sigma_{z}^{\dagger})=0$). This leads to the concept of freezing under \emph{transversal} noise, which we now develop within the setting of metrology.

We first consider a simple model for noisy phase estimation with GHZ states, where bit-flip noise is assumed to act independently and before the parameter imprinting unitary. In this setting, the quantum fisher information (QFI), a figure of merit in quantum metrology, can be calculated directly through quantum state tomography using the experimental setup in Fig.~\ref{Fig:1}. Suppose the unitary imprinting the unknown phase $\varphi$ to be $\exp\{-i \varphi H\}$ with a N-qubit Hamiltonian $H=\sum_{k=1}^N \sigma_i^k$, when $i=x\ (z)$, the unitary acts in a parallel (transversal) fashion to the bit-flip noise.

Figure~\ref{Fig:2}(d) shows how the QFI depends on the noise strength for the noisy 4-qubit GHZ states of Fig.~\ref{Fig:1} when the noise is transversal (yellow and blue lines) and parallel (red and purple lines) relative to the parameter imprinting unitary. Interestingly, the noise dependence of the QFI is different from that of coherence. With transversal noise, the QFI is only constant for the GHZ state prepared in the $\pm$ basis (yellow line), however it performs equally to the SQL corresponding to  using an optimal uncorrelated probe $|++++\rangle$. Instead, for the GHZ state prepared in the $0/1$ basis, the QFI (blue line) does not remain constant and decays under noise, yet always exceeds the SQL.
We see from this simple model that the best scenario for quantum metrology is to use a GHZ state prepared in a basis that coincides with the eigenbasis of the imprinting unitary, and moreover such that the noise is in a transversal direction to guarantee above-SQL performance. It is interesting to point out that coherence and QFI give a different ordering for such two initial states, showing that the coherence resource required for metrology relies on the means of encoding~\cite{Study3}.

{\it Noisy metrology.---}We now consider a more realistic scenario in which the noise occurs simultaneously with the imprinting unitary. In this setting, we focus on estimation of a frequency $\omega$ imprinted as $\exp\{-i\omega t H\}$ with an $N$-qubit Hamiltonian $H=\frac{1}{2}\sum_{k=1}^N \sigma_z^k$.
Following Refs.~\cite{noisymetrology1,noisymetrology2}, we consider a dynamical evolution determined by a Lindblad-type master equation
\begin{equation}
\label{eq}
\frac{\partial \rho(t)}{\partial t}=\mathcal{H}(\rho)+\mathcal{L}(\rho),
\end{equation}
where $\mathcal{H}(\rho)=-i \omega [H,\rho]$ captures the unitary evolution and the Liouvillian $\mathcal{L}(\rho)$ describes the noise. We restrict to uncorrelated noise as this is the most likely form in experiments. Hence, $\mathcal{L}=\sum_k\mathcal{L}^k$ with single-qubit terms
\begin{equation}\label{Eq:Liouvillian}
\mathcal{L}^k(\rho)=\mbox{$-\frac{\gamma}{2}(\rho-\alpha_x\sigma_x^k\rho\sigma_x^k-\alpha_y\sigma_y^k\rho\sigma_y^k-\alpha_z\sigma_z^k\rho\sigma_z^k)$},
\end{equation}
where $\gamma$ is the noise strength (identical for each qubit) and $\alpha_i\geq 0$ with $\alpha_x+\alpha_y+\alpha_z=1$.
When $\alpha_z=1 (\alpha_x=1)$, the noise is parallel (transversal) with respect to the unitary.

The precision in frequency estimation can depend on both the number of qubits $N$ of the probe and the interaction time $t$. In the noise-free setting (i.e., $\gamma = 0$), both quantities can be increased to improve precision~\cite{Entanglementfree}. Increasing $t$ follows the intuitive notion that a longer interaction allows more information to be imparted upon the probe. However, the addition of noise causes the probe state to also deteriorate with time. This trade-off can result in an intermediate-time interaction being optimal for metrology.

We provide here an experimental verification that quantum probes can maintain super-classical performance in metrology despite the presence of noise, focusing on frequency estimation with transversal noise and interrogation-time optimization~\cite{noisymetrology1,noisymetrology2}. Our experimental setup remains as in Fig.~\ref{Fig:1}. We use the $N$-qubit GHZ state $|G_N\rangle=\frac{1}{\sqrt{2}}(|0\rangle^{\otimes N}+|1\rangle^{\otimes N})$ in the computational basis as a probe and perform the metrological procedure for $N=1,2,3,4,6$ qubits.
Significant effort was made to prepare high fidelity probes: for the 2-, 4-, and 6-photon GHZ states, the fidelities were measured to be $0.9961\pm0.0002$, $0.9746\pm0.0027$ and $0.9059\pm0.0078$ respectively (see SM.IC~\cite{SM} for more detail).

Transversal (bit-flip) noise corresponds to $\alpha_x=1$ in Eq.~(\ref{Eq:Liouvillian}). To enact this noise experimentally, one can explicitly solve the master equation in Eq.~(\ref{eq}) and obtain a single-qubit map expressed by a set of Kraus operators $\mathcal{E}_\omega(\rho)=\sum_{i=1}^4 p_i K_i \rho K_i^\dag$, where all $K_i$ are single-qubit unitary operations (see SM.IIIA~\cite{SM} for detail). We can simulate the composite channel $\mathcal{E}_\omega^{\otimes N}$ by letting each photon pass through a HWP sandwiched by two QWPs. This combination can realize an arbitrary single-qubit unitary operation~\cite{QHQ}.
By randomly switching among the angle settings of the waveplates, we can realize each Kraus operator with desired probability (see SM.IIIB~\cite{SM} for detail). Note that the random nature guarantees the Markov property of the simulated noise channel.
In the experiment, we chose an identical noise and signal strength $\omega = \gamma = 1$. Note that when the value $\gamma/\omega$ becomes larger, we require higher preparation fidelity of the GHZ probes to beat the SQL. To characterize the channel, we perform single-qubit process tomography with different evolution times (see SM.IIIB~\cite{SM}
for detail). The average process fidelity is measured to be $0.9952\pm0.0001$, confirming the reliable simulation of the channel.

The frequency $\omega$ is estimated based on the average of measuring a parity operator $P_x=\bigotimes_{k=1}^N \sigma_x^k$, which is optimal for GHZ probes in the noiseless case. The mean-squared error of $\omega$ can be deduced from error propagation,
$\Delta^2\omega=\frac{(\Delta^2 P_x)_{\rho}}{|\partial \langle P_x\rangle_{\rho}/\partial \omega|^2}.$
Since $P_x^2=1$, it follows that $\Delta^2 P_x=1-\langle P_x\rangle^2$. If the measurement is repeated $\nu$ times within a total time $T=\nu t$, the mean-squared error is reduced, as $\Delta^2\omega$ is inversely proportional to $\nu$. We consider the performance of frequency estimation with respect to the total time $T$, and can hence write
\begin{equation}
\label{vaopernorm}
\Delta^2\omega T=t \frac{1- \langle P_x\rangle_t^2}{|\partial \langle P_x\rangle_t/\partial \omega|^2},
\end{equation}
which is valid for the  $T\gg t$ regime. As discussed above, there exists an optimized interrogation time $t^{opt}$ to minimize such a quantity, which we determine by theoretical calculation.

\begin{figure}[tb]
\centering
\includegraphics[width=0.48\textwidth]{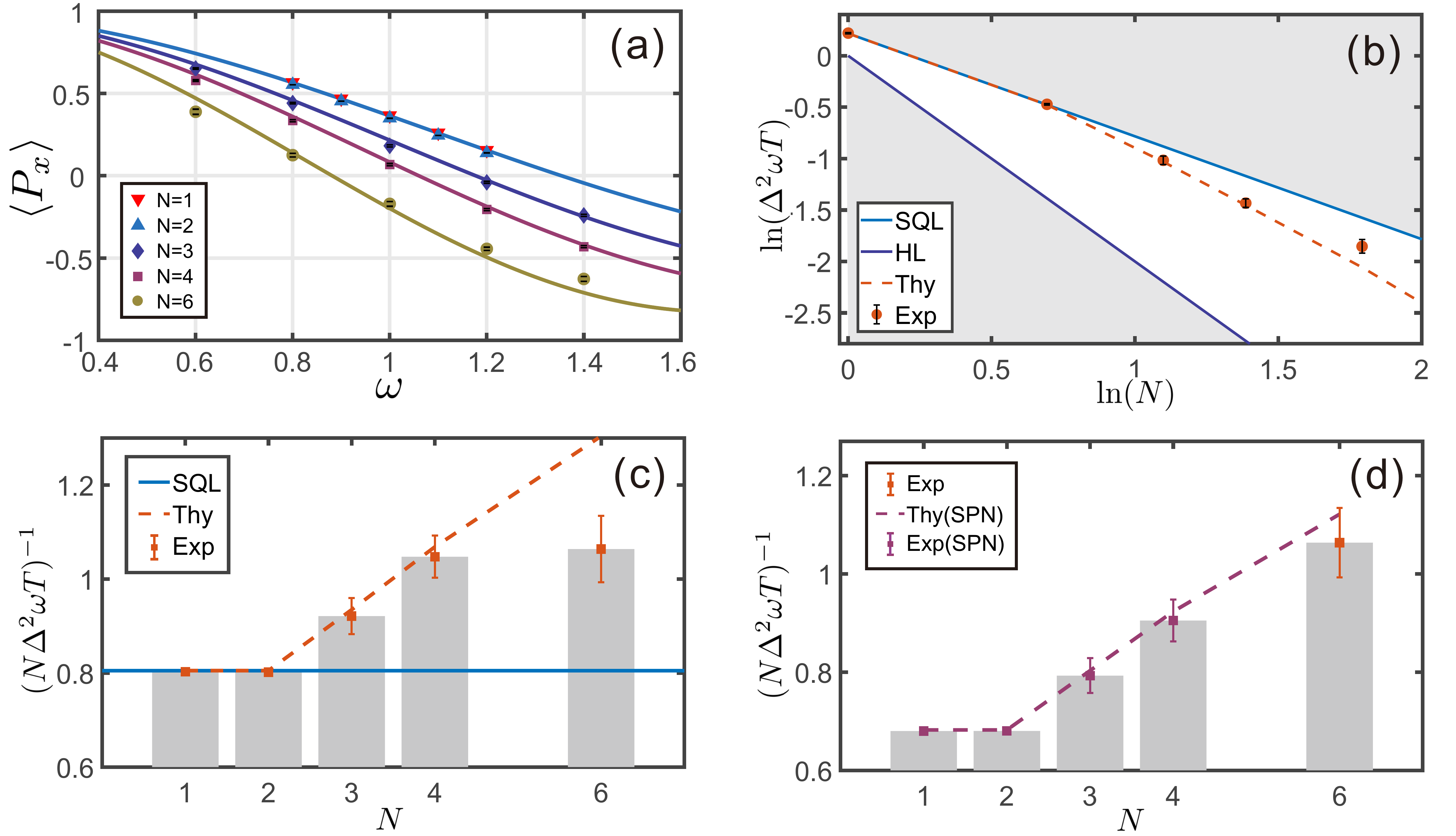}
\vspace*{-.3cm}
\caption{\label{Fig:3} (a) The measured average values of the parity operator $P_x$ for different probe size $N=\{1,2,3,4,6\}$. The optimal evolution times are $t^{opt}=\{1.5490, 0.7745, 0.5684, 0.4650, 0.3508\}$, respectively. The solid lines show theoretical calculations using ideal probes and channels. Note that the red line ($N=1$) is overlapped with the blue line ($N=2$). The error bars are smaller than marker size.  (b) The log-log plot of the mean-squared error of $\omega$ determined by Eq.~\ref{vaopernorm} as a function of the probe size. The results (red dots) are all sandwiched by the SQL and HL bounds, indicating an intermediate scaling. The red dashed line shows the theoretical predictions with ideal probes and channels. (c) Fisher information per photon as a function of $N$ (additional bars are drawn to aid illustration). (d) Fisher information per photon after adding $7\%$ white noise in the $N=1-4$ GHZ probes. \vspace*{-.3cm}}  \end{figure}

It is challenging to measure the partial derivative $\frac{\partial\langle P_x\rangle_t}{\partial \omega}|_{\omega=1}$ experimentally. Here we use the five-point stencil method~\cite{fivestencil},
to approximate the derivative (see SM.IIIC~\cite{SM} for detail). For each $N$, we measured the average values of $P_x$ for five perturbations around $\omega = 1$ with a spacing step of 0.1 or 0.2; the results are summarized in Fig.~\ref{Fig:3}(a).

The mean-squared error of $\omega$ can then be calculated according to Eq.~(\ref{vaopernorm}). Fig.~\ref{Fig:3}(b) shows the log-log plot of the mean-squared error as a function of the probe size $N$, from which we can see that a superclassical precision is demonstrated with GHZ probes of up to six qubits. Note that for parallel noise, there is no benefit in using a highly entangled GHZ state (see SM.IIIA~\cite{SM} for detail). Fig.~\ref{Fig:3}(c) shows a plot of the Fisher information per photon/qubit (defined as $S_N=(N\Delta^2\omega T)^{-1}$) as a function of $N$, from which we can see for high quality GHZ probes ($N=1-4$, fidelity $>0.97$) that the quantity increases with the entangled probe size. Instead, the Fisher information per qubit remains constant for unentangled probes. Note that the 2-photon Bell state performs equivalently to unentangled probes due to suboptimality of the parity measurement under bit-flip noise. For $N=6$, the results do not show that $S_6$ is larger than $S_4$ when taking into account uncertainties, mainly due to the relatively low fidelity of the initial 6-photon GHZ state.
To compare them with similar fidelity, we added further state preparation noise (SPN) in the $N=1-4$ GHZ probes. We consider white noise for simplicity, as it commutes with the incoherent channels. Our procedure was inspired by the effect of white noise on the observed statistics; we randomly flipped the measurement results of the parity operator with probability $p=(1-v_{add})/2$ to simulate adding further white noise $\rho_{add}=v_{add}\rho+(1-v_{add})\frac{\mathbb{I}}{2^N}$~\cite{addnoise}. According to the fidelity difference between $|G_4\rangle$ and $|G_6\rangle$, we set $v_{add}=0.93$ for $N=1-4$. After adding SPN, we observe the expected increasing with $N$ of the Fisher information per qubit (Fig.~\ref{Fig:3}(d)).
In SM.IIID~\cite{SM}, we give a numerical analysis showing that a fixed amount of noise in the initial probes does not impact how the metrological precision scales, while preparation noise that increases with $N$ can prevent superclassical scaling; if we treat such scaling preparation noise as a deviation from perfectly transversal noise in the channel, the result is similar to that discussed in~\cite{noisymetrology1}. In the SM.IIIE~\cite{SM}, we also determine the mean-squared error of $\omega$ by using the QFI of the evolved state $\mathcal{F}(\rho_\omega)$ (N=1-4), which also shows a superclassical precision scaling.

In conclusion, we have experimentally investigated the resilient effect of quantum coherence and QFI under transversal noise. In particular, we demonstrated that both quantities can be fully protected during the evolution in a 4-photon GHZ state. We also investigated such resilience in a realistic metrology task where noise and parameter imprinting occur simultaneously. By configuring the signal Hamiltonian perpendicular to the noise, we demonstrated that our prepared GHZ probes can beat the SQL with up to 6-photon probes. For very pure probes (up to 4-photon), our results show that a quantum advantage in metrological scaling can survive even in the presence of uncorrelated Markovian noise, paving the way for scalable noisy metrology with only passive noise control.  Future perspectives can be to combine our analysis with active error correction in metrology \cite{QECmetrology1,QECmetrology2}. It would also be extremely attractive to find other applications which can harness the natural resilient effect of quantum resources against decoherence, especially in quantum computation.

{\it Acknowledgements.---}This work was supported by the National Key Research and Development Program of China (Grant No.~2017YFA0304100), the National Natural Science Foundation of China (Grants No.~61327901, No.~11774335, No.~11734015, No.~11704371, and No.~11821404), Key Research Program of Frontier Sciences, CAS (Grant No.~QYZDY-SSW-SLH003), the Fundamental Research Funds for the Central Universities (Grants No.~WK2470000026 and No.~WK2470000018), Anhui Initiative in Quantum Information Technologies (Grants No.~AHY020100 and No.~AHY070000), Science Foundation of the CAS (Grant No.~ZDRW-XH-2019-1), China Postdoctoral Science Foundation (Grant No.~2017M612074), and the European Research Council (Starting Grant GQCOP No.~637352).

{\it Note added.---}After completion of this work, we became aware of a related work~\cite{local}, which demonstrates that local encoding provides a practical advantage for phase estimation in noisy environments. \nocite{CRbound,QFI} 

\vfill

\cleardoublepage

\begin{widetext}

\appendix*
\renewcommand{\thesection}{SM.\Roman{section}}
\renewcommand{\thesubsection}{\Alph{subsection}}
\renewcommand{\theequation}{S\arabic{equation}}
\renewcommand{\thefigure}{S\arabic{figure}}
\setcounter{equation}{0}
\setcounter{figure}{0}

\section*{APPENDIX: SUPPLEMENTAL MATERIAL}

\section{Experimental details}

\subsection{EPR source}

The detailed structure of the EPR source is shown in Fig.\ref{Fig:S1}. The two $\beta$-barium borate (BBO) crystals are both 1-mm thick and have the same cutting angles for beamlike type-II phase-matching, a true-zero-order half-wave plate (THWP) is inserted between them. In beamlike SPDC emission, the down-converted photon pairs are emitted into two separate beams with a Gaussian-like intensity distribution, which will benefit for achieving a high collection efficiency and high counting rate. In the sandwich-like structure, the polarization state of the photon pairs generated by BBO1 can be written as $|H\rangle_{e,2}|V\rangle_{o,1}$, where H (V) denotes horizontal (vertical) polarization, the subscript o (e) denotes the ordinary (extraordinary) photon with respect to the BBO crystal, and subscript 1, 2 denote different spatial modes. After transmitting the THWP, the polarization state rotates to $|V\rangle_{e,2}|H\rangle_{o,1}$. BBO2 is placed in the same manner as BBO1, thus generate photon pairs also in the state $|H\rangle_{e,2}|V\rangle_{o,1}$. The two possible ways of generating photon pairs are further made indistinguishable by carefully spatial (LiNbO$_3$ crystals) and temporal (YVO$_4$ crystals) compensations, thus the photon pairs are prepared in the state of  $(|H\rangle_{e,2}|V\rangle_{o,1}+|V\rangle_{e,2}|H\rangle_{o,1})/\sqrt{2}$. Then a HWP is used to transform this state to $(|HH\rangle+|VV\rangle)/\sqrt{2}$, which is required for GHZ state preparation. Note that the sandwich structure generate the e- and o-photons into different spatial modes, which meets the key requirement of the entanglement concentration scheme and allows us to engineer different narrowband filters for e- and o-photons.

\begin{figure}[b!]
\centering
\includegraphics[width=0.6\textwidth]{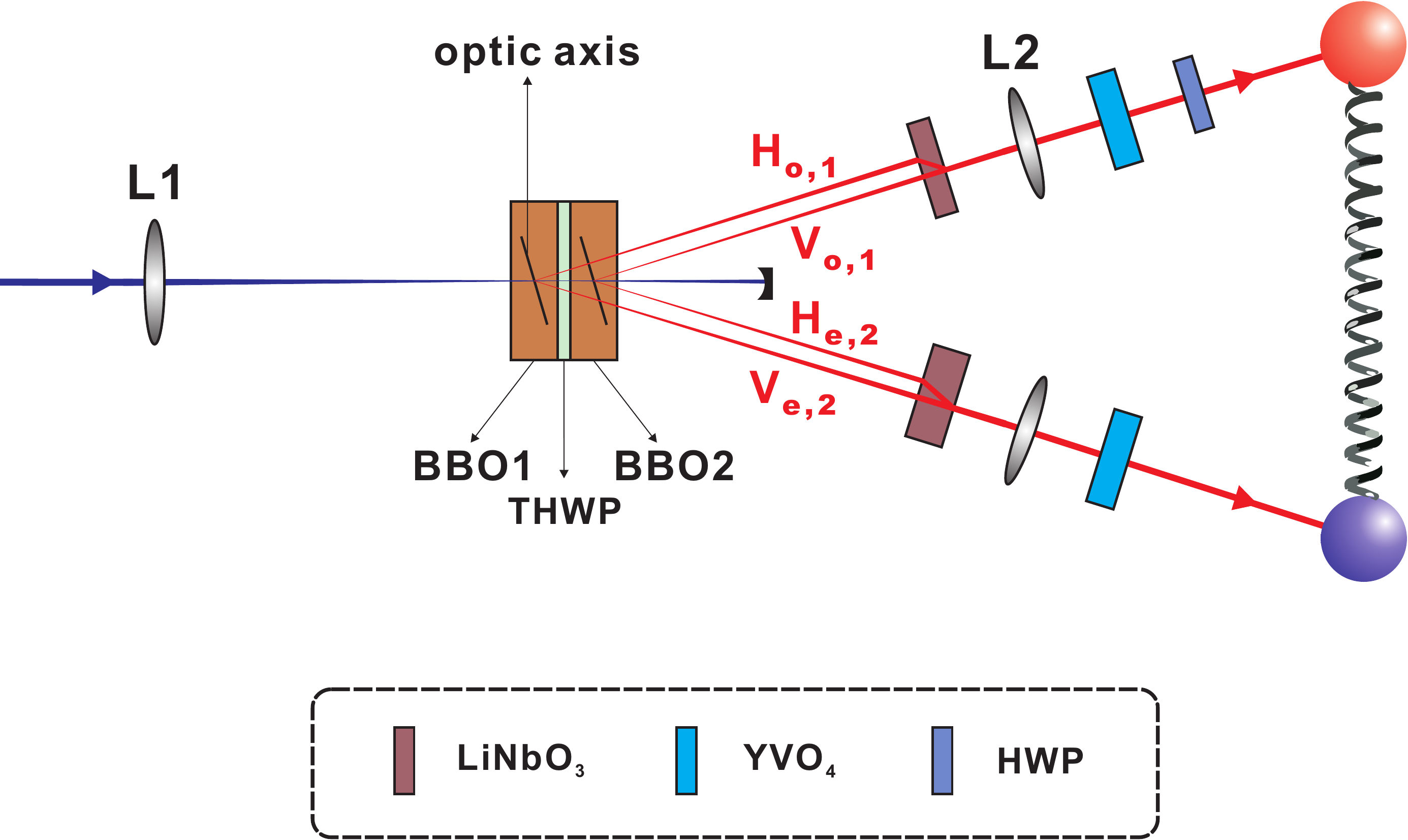}
\caption{\label{Fig:S1} Sandwich-like EPR source.}  \end{figure}

\subsection{GHZ state preparation}

A high-accuracy multiphoton interferometer is crucial to generate high fidelity multiphoton entangled states. Fig.\ref{Fig:S2} shows the optical network to generate the 6-photon GHZ state. The entangled photon pairs are generated from the above EPR sources. The three e-photons (the blue balls) are directed to overlap on two polarization beam-splitters (PBSs). The PBSs are set to transmit (reflect) H- (V-) polarized photons. After the interferometer, the H- and V-polarized photons will be redistributed among the output modes $e_1'$, $e_2'$ and $e_3'$. When there is one and only one photon in each output port, the PBS acts as a parity check operator $|HH\rangle \langle HH|+ |VV\rangle \langle VV|$ between the two input photons, which is widely used to fuse independent photon pairs. One can check that after the two PBSs only two possible terms $|H\rangle^{\otimes 6}$ and $|V\rangle^{\otimes 6}$ are post-selected. The two terms are further made indistinguishable carefully, thus the input EPR photon pairs are projected into the 6-photon GHZ state. The time delays between different paths are finely adjusted by using P1 and P2 to ensure that different polarized photons arrive at each output port simultaneously. We insert a quartz plate (QP) before the output port $e_1'$ to compensate the small time difference. Then all photons pass through narrowband filters and single mode fibers for spectral and spatial selection. In addition, we insert a sequence of wave plates which consists of one HWP sandwiched by two $45^\circ$ QWP before one output port to carefully calibrate the relative phase between the two terms in the GHZ state. By removing one EPR source and the corresponding PBS, we can generate the 4-photon GHZ state.

\begin{figure}[tb]
\centering
\includegraphics[width=0.7\textwidth]{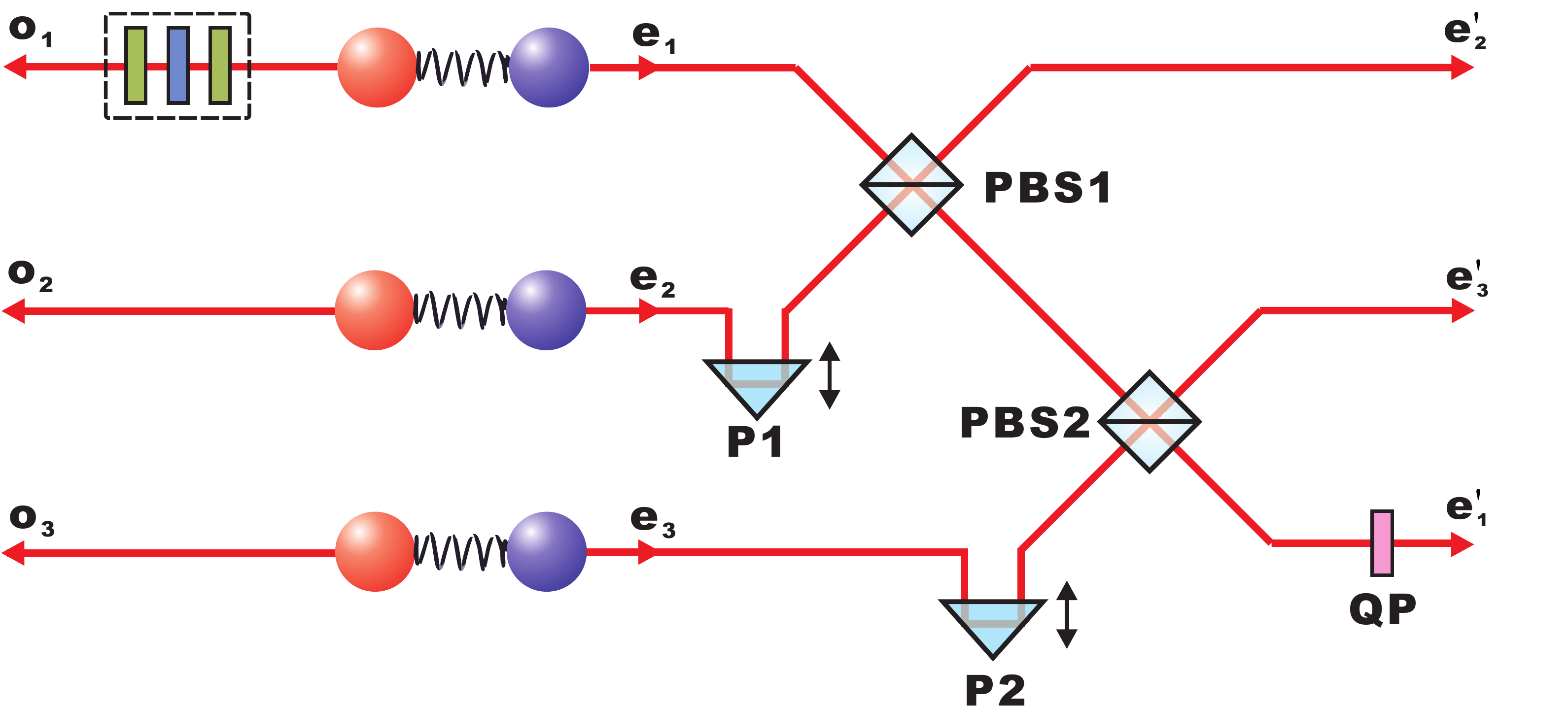}
\caption{\label{Fig:S2} The optical network to generate the 6-photon GHZ state. }  \end{figure}

\subsection{Characterizing the prepared GHZ states}

To characterize the prepared N-photon GHZ states, we measure the state fidelity $F_{G_N}=\langle G_N|\rho_{exp} |G_N\rangle$ of them, where $|G_N\rangle=\frac{1}{\sqrt{2}}(|H\rangle^{\otimes N}+|V\rangle^{\otimes N})$ denotes the ideal N-qubit GHZ state. We measure the projection operator
\begin{equation}
\label{witness}
\left\vert G_{6}\right\rangle\left\langle G_{6}\right\vert=\frac{1}{2}A+\frac{1}{12}\overset{5}{\underset{k=0}{\sum }}(-1)^{k}M _{k},
\end{equation}
where $A=\left\vert H\right\rangle ^{\otimes 6}\left\langle H\right\vert^{\otimes 6}+\left\vert V\right\rangle ^{\otimes 6}\left\langle V\right\vert^{\otimes 6}$ and $M _{k}=[\cos (\frac{k\pi }{6})\sigma _{x}+\sin (\frac{k\pi }{6})\sigma _{y}]^{\otimes 6}$ for N=6, and do full state tomography for N=2 and N=4. The experimental results are summarized in Fig.\ref{Fig:S3}. The fidelities are measured to be $0.9961\pm0.0002$, $0.9746\pm0.0027$ and $0.9059\pm0.0078$ for the 2-, 4-, and 6-photon GHZ state respectively, the counting rates are 10000 Hz, 1 Hz and 0.04 Hz respectively. In the six photon experiment we use 8- and 3-nm bandwidths filters for o- and e-photons respectively, while in the two and four photon experiment we use 3- and 2-nm bandwidths filters for o- and e-photons respectively. When the filter setting changes from $8 \ \&\  3 nm$ to $3 \ \&\  2 nm$, we observe the Hong-Ou-Mandel interference visibility increased from 0.92 to 0.97, while the collection efficiency of the photon source drop from 0.3 to 0.2.

\begin{figure}[tb]
\centering
\includegraphics[width=0.7\textwidth]{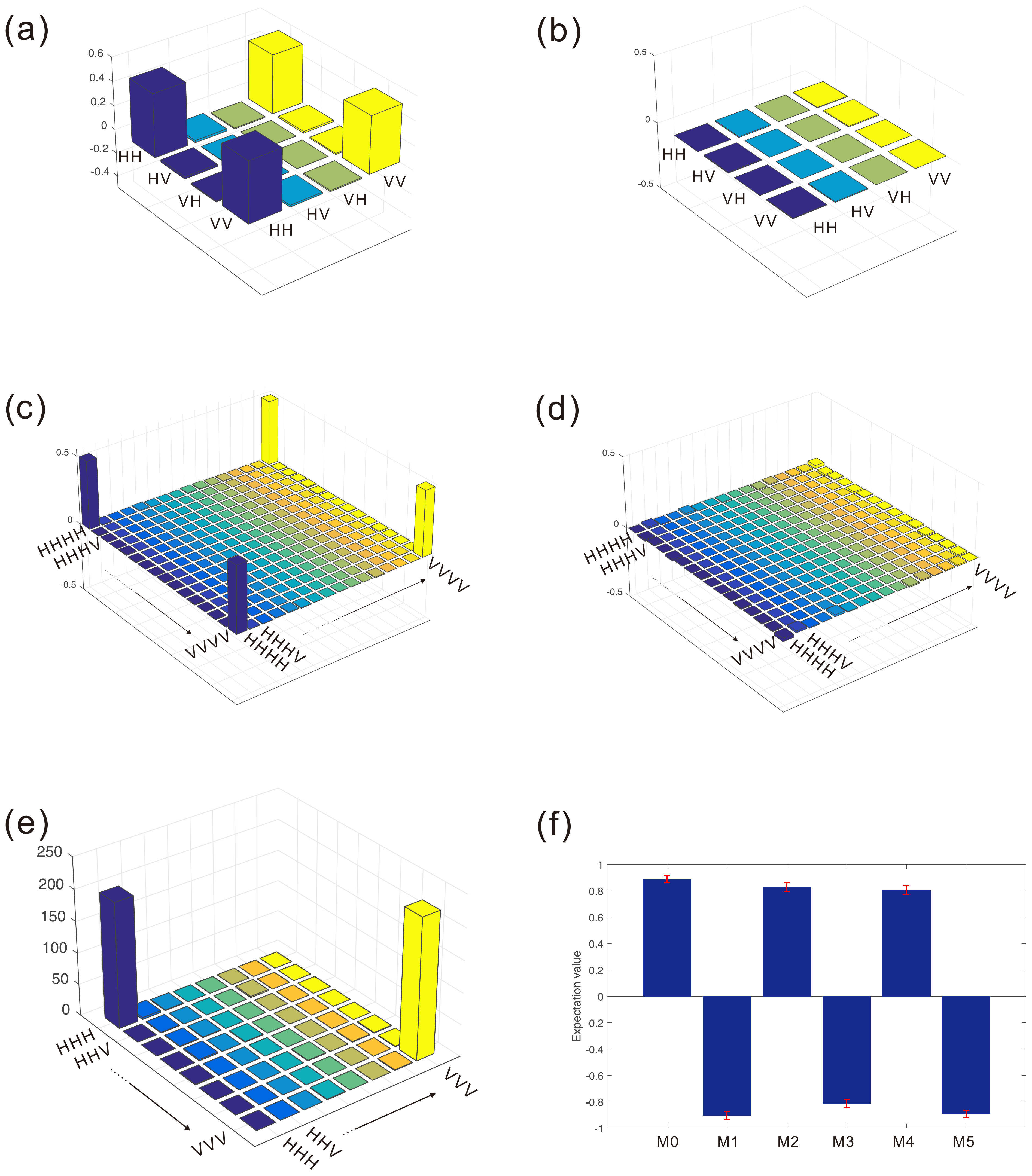}
\caption{\label{Fig:S3} (a) Real and (b) imaginary part of the experimentally reconstructed density matrix of the 2-photon Bell state. (c) Real and (d) imaginary part of the reconstructed 4-photon GHZ state. (e) Results for A measurement. (f) Expectation values of $M_k (k=0,\cdots,5)$.  }  \end{figure}

\section{Frozen quantum coherence}

\subsection{Quantum coherence quantifiers}

The resource framework of quantum coherence is based on identifying a set of incoherent states and a class of incoherent operations that map the set onto itself, which is analog to the resource theory of entanglement. For a fixed reference basis $\{|i\rangle\}$, the incoherent states are defined as diagonal states in this basis $\delta=\sum_i p_i |i\rangle\langle i|$, where $p_i$ are probabilities. We denote the set of incoherent states as $\mathcal{I}$.

A completely positive trace-preserving (CPTP) map $\Lambda$ can be characterized by a set of Kraus operators $\Lambda(\rho)=\sum_j K_j \rho K_j^\dag$, where $\sum_j K_j^\dag K_j=\mathbb{I}$. Then incoherent operations or channels (ICPTP maps) satisfy the additional constraint $K_j \mathcal{I} K_j^\dag \subset \mathcal{I}$ for all j, which are closed to the set of incoherent states.

Baumgratz et al~\cite{frame}. have proposed a set of requirements which should be satisfied by any valid coherence measure $C(\rho)$:

(C1) $C(\rho)\geq0$, and $C(\rho)=0$ if and only if $\rho\in\mathcal{I}$.

(C2) Contractivity under incoherent channels $C(\rho)\geq C(\Lambda_{ICPTP}(\rho))$.

(C3) Contractivity under selective measurements on average, $C(\rho)\geq\sum_j p_j C(\rho_j)$, where $p_j=Tr(K_j\rho K_j^\dag)$ and $\rho_j=K_j\rho K_j^\dag/p_j$, the Kraus operators satisfy that $\sum_j K_j^\dag K_j=\mathbb{I}$ and $K_j\mathcal{I}K_j^\dag\subset\mathcal{I}$ for all j.

(C4) Convexity, $C(q\rho+(1-q)\tau)\leq qC(\rho)+(1-q)C(\tau)$ for any states $\rho$,$\tau$ and $q\in[0,1]$.

A generic distance-based measure of coherence is defined as
\begin{equation}\label{eq:CDdef}
{\cal C}_{D} (\rho) = \min _{\delta \  \in \ \mathcal{I}} D (\rho, \delta) = D (\rho, \delta_{\rho})\,,
\end{equation}
where  $\delta_{\rho}$ is one of the closest incoherent states to $\rho$ with respect to $D$. Notable examples that satisfy all the four properties mentioned above include the $l_1$ norm of coherence, via summing over the off-diagonal elements ${\cal C}_{l_{1}}(\rho) = \sum_{i \neq j} \left| \rho_{ij} \right|$, and the relative entropy of coherence ${\cal C}_{RE} (\rho) = {\cal S}(\rho_{\rm diag}) - {\cal S}(\rho)$, where $\rho_{\rm diag}$ is the matrix containing only the diagonal elements of $\rho$, and ${\cal S}(\rho) = -\text{Tr}(\rho \log \rho)$ is the von Neumann entropy.

\subsection{Simulating the bit-flip noise}
Experimentally, the bit-flip channels are engineered by a randomized setting of the HWP angles to be $\pm\theta$ with equal probability, and each HWP is sandwiched by two quarter-wave plates (QWPs) at $90^\circ$ with respect to the horizontal direction. One can check that such evolution enacts bit-flip noise $\mathcal{E}(\rho)=(1-p/2)\rho+p\sigma_x\rho\sigma_x$ with the noise strength $p=2\sin^22\theta$.

\subsection{Tomographic results}

In Fig.\ref{Fig:S4}, we give the tomographic results of the evolved four-photon GHZ states with state preparation and reference basis chosen as 0/1 or $\pm$ basis under bit-flip noise, which correspond to the four cases studied in Fig.2 in the main text.

\begin{figure}[tb]
\centering
\includegraphics[width=0.98\textwidth]{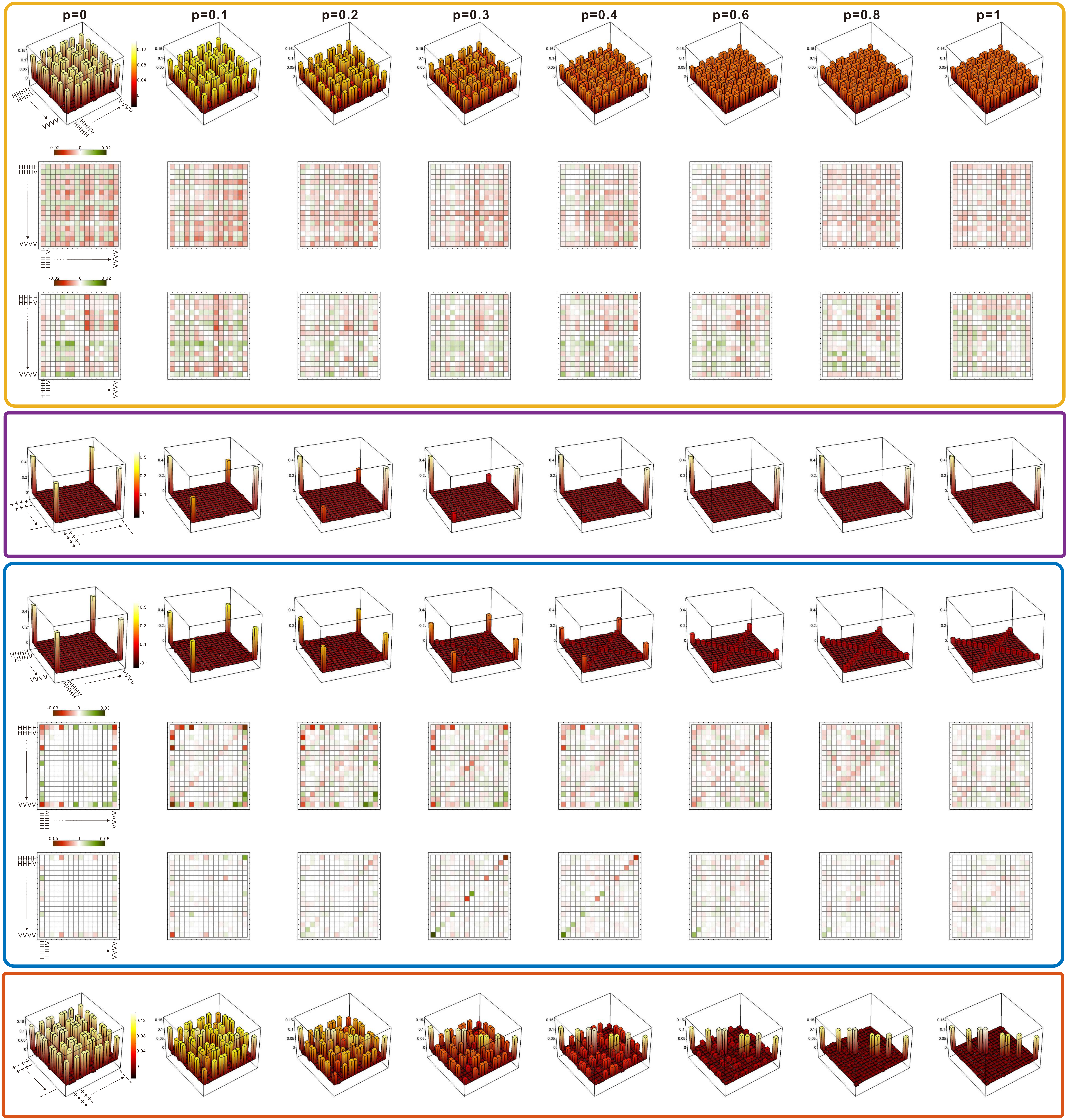}
\caption{\label{Fig:S4} Tomographic results of eight different states during the evolution, the corresponding noise strength are $p=\{0,0.1,0.2,0.3,0.4,0.6,0.8,1\}$. In each box, the 3D bar graphs show the real part of the experimental reconstructed density matrices. The box colors correspond to the colors used in Fig.2 in the main text. In the orange and blue boxes, we also show the difference between the real part (the second line) and the imaginary part (the third line) of the reconstructed density matrices with ideal ones. Note that we mark the color bar and the reference basis only in the first plot in each line.}  \end{figure}

\section{Noisy metrology}

\subsection{Solving the master equation}

According to Ref.~\cite{noisymetrology1}, the single-qubit map of the master equation Eq.~({1}) in the main text for time t can be expressed in the Pauli basis
\begin{equation}
\label{map}
\mathcal{E}_{\omega} \left( \rho \right) = \sum_{i,j} S_{ij} \sigma_{i} \rho \sigma_{j} ,
\end{equation}
where the Pauli operators ${\sigma}_{0}\!\equiv\!\mathbb{I}$, ${\sigma}_{1}\!\equiv\!{\sigma}_{x}$, ${\sigma}_{2}\!\equiv\!{\sigma}_{y}$ and ${\sigma}_{3}\!\equiv\!{\sigma}_{z}$. All elements of the matrix S are zero, except $S_{00} =(a+b)/2$, $S_{11} =(d+f)/2$, $S_{22} =(d-f)/2$, $S_{33} =(a-b)/2$, $S_{03} =ic/2$, $S_{30} =-ic/2$. The coefficients $a,b,c,d$, and $f$ are real and depend on $\omega$, $\gamma$, and $t$:
\begin{align}
\label{parameter}
a  & =\frac{1}{2}e^{-\frac{1}{2}t\left(  1+\alpha_{x}+\alpha_{y}-\alpha
_{z}\right)  \gamma}\left(  1+e^{t\left(  \alpha_{x}+\alpha_{y}\right)
\gamma}\right)  \nonumber \\
b  & =\frac{1}{2}e^{-\frac{1}{2}t\left(  \gamma+\alpha_{z}\gamma+\sqrt{\left(
\alpha_{x}-\alpha_{y}\right)  ^{2}\gamma^{2}-4\omega^{2}}\right)  }\left(
1+e^{t\sqrt{\left(  \alpha_{x}-\alpha_{y}\right)  ^{2}\gamma^{2}-4\omega^{2}}%
}\right)  \nonumber\\
d  & =\frac{1}{2}e^{-\frac{1}{2}t\left(  1+\alpha_{x}+\alpha_{y}-\alpha
_{z}\right)  \gamma}\left(  -1+e^{t\left(  \alpha_{x}+\alpha_{y}\right)
\gamma}\right)  \\
f  & =\frac{e^{-\frac{1}{2}t\left(  \gamma+\alpha_{z}\gamma+\sqrt{\left(
\alpha_{x}-\alpha_{y}\right)  ^{2}\gamma^{2}-4\omega^{2}}\right)  }\left(
-1+e^{t\sqrt{\left(  \alpha_{x}-\alpha_{y}\right)  ^{2}\gamma^{2}-4\omega^{2}}%
}\right)  \left(  \alpha_{x}-\alpha_{y}\right)  \gamma}{2\sqrt{\left(
\alpha_{x}-\alpha_{y}\right)  ^{2}\gamma^{2}-4\omega^{2}}}\nonumber\\
c  & =\frac{e^{-\frac{1}{2}t\left(  \gamma+\alpha_{z}\gamma+\sqrt{\left(
\alpha_{x}-\alpha_{y}\right)  ^{2}\gamma^{2}-4\omega^{2}}\right)  }\left(
-1+e^{t\sqrt{\left(  \alpha_{x}-\alpha_{y}\right)  ^{2}\gamma^{2}-4\omega^{2}}%
}\right)  \omega}{\sqrt{\left(  \alpha_{x}-\alpha_{y}\right)  ^{2}\gamma
^{2}-4\omega^{2}}}\nonumber
\end{align}

The Kraus representation of the map $\mathcal{E}_{\omega}$ can be obtained by diagonalizing the matrix S, the normalized eigenvectors and the eigenvalues give the Kraus operators and the corresponding probability respectively:
\begin{eqnarray}
\label{Kraus}
\mathcal{E}_{\omega} \left( \rho \right) &=& \sum_{i}^4 p_i K_i \rho K_i^\dag, \nonumber \\
p_1 =\frac{a+\sqrt{b^2+c^2}}{2}, p_2&=&\frac{a-\sqrt{b^2+c^2}}{2}, p_3=\frac{d+f}{2}, p_4=\frac{d-f}{2}\\
K_1 =\begin{pmatrix}
		1 & 0\\
		0 & e^{i\theta}\\
	\end{pmatrix},\ \
	K_2&=&\begin{pmatrix}
		1 & 0\\
		0 & -e^{i\theta}\\
	\end{pmatrix},\ \
    K_3=\sigma_1,\ \
    K_4=\sigma_2. \nonumber
\end{eqnarray}
where $\theta=\arccos \frac{b}{\sqrt{b^2+c^2}}$. Then the composite channel $\mathcal{E}_{\omega}^{\otimes N}$ can be written as
\begin{equation}
\label{mapN}
\mathcal{E}_{\omega}^{\otimes N} \left( \rho \right) = \sum_{i_1 i_2 \cdots i_N}^4 p_{i_1} p_{i_2}\cdots p_{i_N} K_{i_1}K_{i_2}\cdots K_{i_N} \rho K_{i_1}^\dag K_{i_2}^\dag \cdots K_{i_N}^\dag.
\end{equation}

For a GHZ state probe $|G_N\rangle=\frac{1}{\sqrt{2}}(|0\rangle^{\otimes N}+|1\rangle^{\otimes N})$, it is easy to calculate the output state $\rho_\omega$ due to the permutation symmetry of the parties of the input state and the channel. One can find that the evolved state is diagonal in the GHZ basis and an ``X'' state in the computational basis. The diagonal and anti-diagonal terms are given by
\begin{eqnarray}
\label{elements}
\langle m|\rho_\omega|m\rangle &=& \frac{1}{2}\left[(S_{11}+S_{22})^m (S_{00}+S_{33}+S_{03}+S_{30})^{N-m}+(S_{11}+S_{22})^{N-m} (S_{00}+S_{33}+S_{03}+S_{30})^m\right]\nonumber \\
&=&\frac{1}{2}\left[d^m a^{N-m}+d^{N-m} a^m\right],\\
\langle m|\rho_\omega|m'\rangle &=& \frac{1}{2}\left[(S_{11}-S_{22})^m (S_{00}-S_{33}-S_{03}+S_{30})^{N-m}+(S_{11}-S_{22})^{N-m} (S_{00}-S_{33}+S_{03}-S_{30})^m\right]\nonumber \\
&=&\frac{1}{2}\left[f^m (b-ic)^{N-m}+f^{N-m} (b+ic)^m\right].\nonumber
\end{eqnarray}
where $m$ denotes the number of qubit 1 in the state vector, while $m'=N-m$. Each m-term has a degeneracy of $\binom{N}{m}$. The average value of the parity operator can be obtained by sum over the anti-diagonal terms
\begin{equation}
\label{Px}
\langle P_x\rangle_{\rho_\omega}=\frac{1}{2}\left[(f+b-ic)^N+(f+b+ic)^N\right].
\end{equation}
Also the diagonalization of $\rho_\omega$ is reduced to the diagonalization of $\left\lfloor N/2 +1\right\rfloor$ different $2 \times2 $ density matrices, which simplifies the calculation dramatically.

For parallel noise ($\alpha_x=0, \alpha_y=0, \alpha_z=1$), we see $d=f=0$ and $a=1$, the evolved GHZ state $\rho_\omega$ only has four terms and only the two anti-diagonal terms are dephasing. It's easy to calculate the quantum fisher information of the state $\mathcal{F}(\rho_\omega)=N^2t^2e^{-2N\gamma t}$, thus the mean-squared error of $\omega$ is bounded by $\Delta^2\omega T\geq\sqrt{\frac{2\gamma e}{N}}$ with optimized interrogation-time $t^{opt}=\frac{1}{2N\gamma}$. However, the result is exactly equal to the SQL bound. In this case the parity measurement is optimal and can deduce the same precision~\cite{parallel}. Thus a GHZ strategy is useless for parallel noise. It was further proved that even optimizing the input state, the asymptotic scaling is still SQL-like and the quantum strategies provide only a constant factor improvement of $\sqrt{e}$~\cite{nogo1}.

For transversal noise ($\alpha_x=1, \alpha_y=0, \alpha_z=0$), while we do not have an analytical expression, we can efficiently compute the optimized interrogation-time and determine the mean-squared error numerically by using the parity measurement or quantum fisher information method. Although the parity measurement is suboptimal in this scenario, as shown in~\cite{noisymetrology2}, one can prove the superclassical precision scaling of $1/N^{5/3}$ for both methods by taking $t\propto 1/N^{1/3}$.

\subsection{Simulating the unitary+noise channel}

The channel is simulated by one HWP sandwiched by two QWPs, see Fig.1(b) in the main text. According to the Kraus decomposition Eq.\ref{Kraus}, the settings of these plates to realize the four Kraus operators are $(\frac{\pi}{4},\frac{\theta+\pi}{4},\frac{\pi}{4})$, $(\frac{\pi}{4},\frac{\theta+2\pi}{4},\frac{\pi}{4})$, $(\frac{\pi}{4},\frac{\pi}{4},\frac{3\pi}{4})$, and $(\frac{\pi}{4},\frac{\pi}{2},\frac{3\pi}{4})$, respectively. To randomly switch between the settings, all the wave plates are mounted on motorized  rotation stages and controlled by a QRNG from ID Quantique. Each time the QRND generate a random number $r$ between 0 and 1: if $r\in[0,p_1)$ we set the wave plates to realize $K_1$; else if $r\in[p_1,p_1+p_2)$ we set the wave plates to realize $K_2$, and so on. In the 3-, 4- and 6-photon experiments we set the wave plates to refresh every second (exclude the rotation time of the motors). The counting rates for 3-, 4- and 6-photon experiments are 0.5 Hz, 1 Hz and 0.04 Hz respectively, thus approximately every copy experiences a random switch among the settings. In the 1- and 2-photon experiment, we simply perform each Kraus term in Eq.\ref{mapN} with a duration proportional to its probability during data collection, from a statistical point of view the result is equal to using a random switch.

To characterize the channel, we perform single-qubit process tomography with different evolution time t=0.5, 1, and 1.5 for the parameters $\gamma=\omega=1, \alpha_x=1, \alpha_y=0, \alpha_z=0$. The reconstructed matrices S in Eq.\ref{map} are shown in Fig.\ref{Fig:S5}. The process fidelities are calculated to be $0.9922\pm0.0001$, $0.9948\pm0.0001$, and $0.9986\pm0.0001$ respectively, testifying the credible simulation of the channel.
\begin{figure}[tb]
\centering
\includegraphics[width=0.97\textwidth]{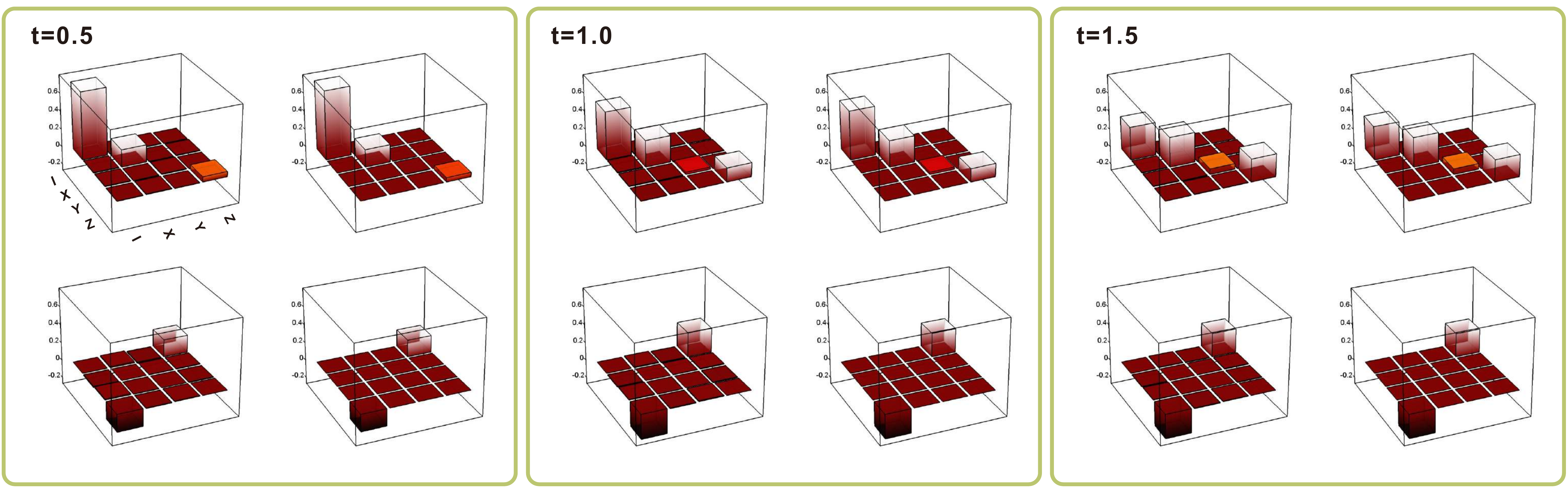}
\caption{\label{Fig:S5} Tomographic results of the channel in Pauli basis. In each box, the left two pictures show the real (top) and imaginary (bottom) part of the experimental reconstructed density matrix, the right pictures show the theoretical density matrix. }  \end{figure}

\subsection{Five-point stencil}

The first derivative of a function $f$ at a point $x$ can be approximated by using a five-point stencil method
\begin{equation}
f'(x)\approx \frac{-f(x+2h)+8f(x+h)-8f(x-h)+f(x-2h)}{12h}.
\end{equation}
where $h$ is the grid length. The error of this approximation is of the order $h^4$, which can be obtained from Taylor expansion of the right-hand side
\begin{equation}
\frac{-f(x+2h)+8f(x+h)-8f(x-h)+f(x-2h)}{12h}=f'(x)-\frac{1}{30}f^{(5)}(x)h^4+O(h^5).
\end{equation}
In the experiment, we use such method to determine the partial derivatives. Although it is better to use small h to reduce the estimated error, we find that if $h$ is too small the photon statistics will contribute a very large error bar to the results. Thus we choose $h=0.1$ or 0.2 in the experiment. In Table.S1 we compare the contribution of these two errors to the final results, from which we can see the estimated error is much smaller than the photon statistical error.

\begin{table}[htb]
\centering
\caption{The results of mean-squared error of $\omega$ deduced from parity measurement or quantum fisher information with different probe size. Error1, the error due to photon statistics. Error2, the error due to five-point stencil. In the experiment, we choose the grid length h=0.1 for N=1,2 and h=0.2 for N=3,4,6. Error2 are always negative which means the five-point stencil method can give a lower bound of the partial derivatives. }
\begin{tabular}{|c|c|c|c|c|c|} \hline
 N & 1 & 2 & 3 & 4 & 6 \\\hline
$(\Delta^2\omega T)_{P_x}$ & 1.2461 & 0.6237 & 0.3619 & 0.2387 & 0.1567\\\hline
Error1 & $\pm0.0066$ & $\pm0.0043$ & $\pm0.0150$ & $\pm0.0102$ & $\pm0.0104$\\\hline
Error2 & -0.00004 & -0.00002 & -0.00024 & -0.00019 & -0.00018\\\hline
$(\Delta^2\omega T)_{QFI}$ & 1.1925 & 0.5487 & 0.3338 & 0.1866 & \\\hline
Error1 & $\pm0.0003$ & $\pm0.0062$ & $\pm0.0286$ & $\pm0.0364$ & \\\hline
Error2 & -0.00005 & -0.00004 & -0.00064 & -0.00072 &\\\hline
\end{tabular}
\end{table}

\subsection{Numerical analysis with SPN}

We analyze a more realistic scenario in which the fidelity of the initial GHZ states decreases with the qubit number, e.g., it is reasonable to consider each additional qubit along with an entangling gate contributes to the same amount of noise in the GHZ state preparation circuit. Thus the N-qubit GHZ state's fidelity is equal to $(F_0)^N$, where $F_0$ is the qubit-number-normalized fidelity. If we consider  white noise for simplicity $\rho_{noise}=v|G_N\rangle\langle G_N|+(1-v)\frac{\mathbb{I}}{2^N}$, the visibility $v$ is approximately equal to the fidelity for large N. Fig.\ref{Fig:S6} shows the numerical analysis of the precision scaling determined by parity measurement with different values of $F_0$. From which we can see any amount of scaling preparation noise can destroy the superclassical precision when N is large enough. However, the SQL bound can be surpassed when N is small, thus demonstrating the benefit of the transversal configuration of the channel. In addition, we show from the red line that a fixed amount of initial noise doesn't impact on the precision scaling.

\begin{figure}[tb]
\centering
\includegraphics[width=0.5\textwidth]{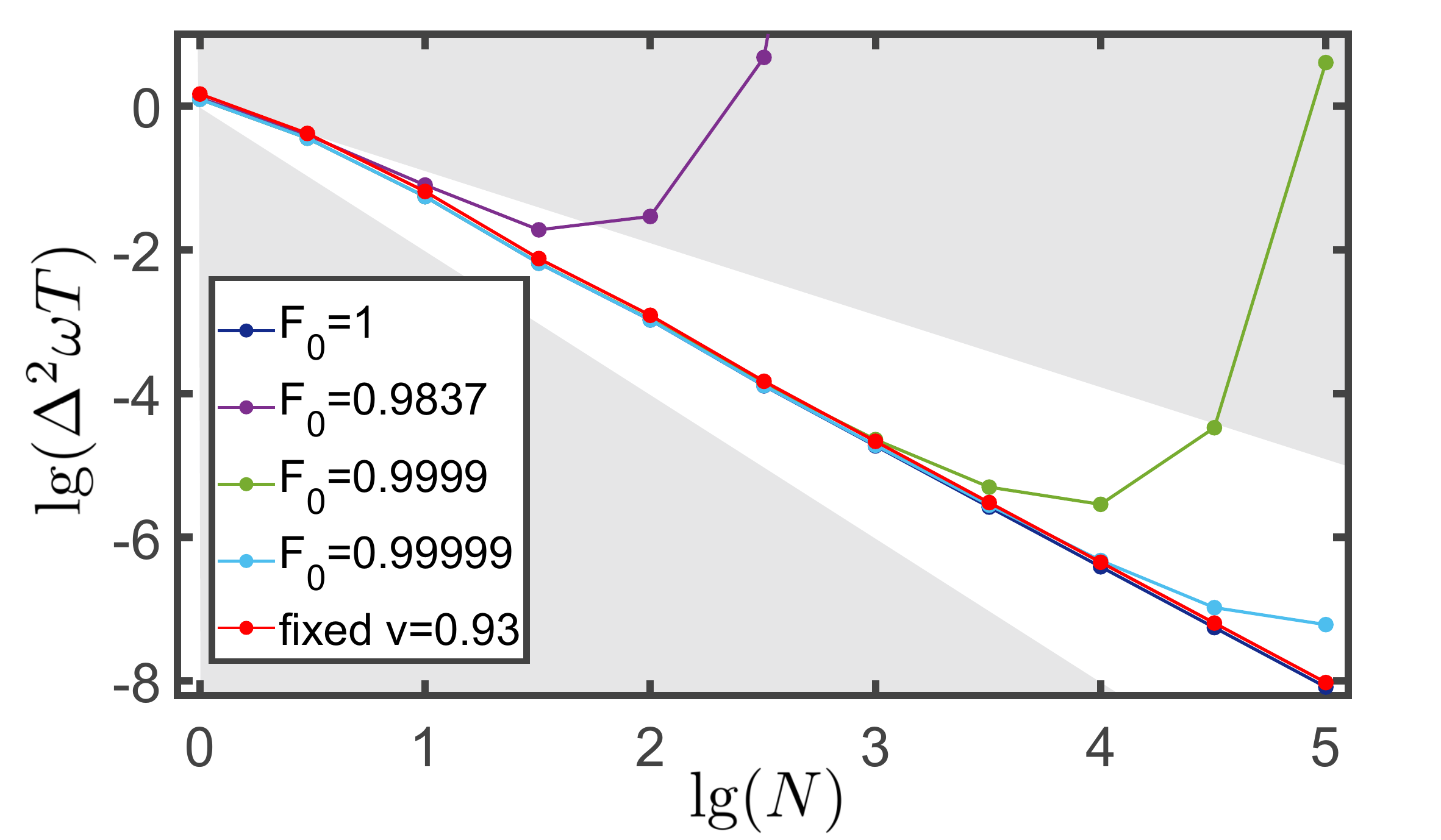}
\caption{\label{Fig:S6} Numerical analysis of the precision scaling with white noise in the initial probes with N up to $10^5$. The above and below shaded areas represent the SQL and HL respectively. The dark blue, purple, green and light blue lines show the precision scaling for $F_0=\{1,0.9837,0.9999,0.99999\}$ respectively. $F_0=0.9837$ is the fidelity level in our six-photon experiment. The red line shows the precision scaling with a fixed $7\%$ white noise in the initial probes.}  \end{figure}

\subsection{Determining the mean-squared error of $\omega$ by QFI}

We give the tomographic results of the evolved N-photon GHZ states in the noisy metrology experiment, as shown in Fig.\ref{Fig:S7}-\ref{Fig:S10}. For N=1,2,3,4, the optimal interrogation times are calculated to be $t^{opt}=\{1.7006, 0.9498, 0.7692, 0.6821\}$ respectively. The precision scaling determined by quantum fisher information is shown in Fig.\ref{Fig:S11}, according to the quantum Crame\'r-Rao bound $\Delta^2 \omega T\geq 1/(\mathcal{F}(\rho_\omega)/t)$~\cite{CRbound}. The quantum fisher information quantifies the largest information extractable from the probe, which is defined as~\cite{QFI} $\mathcal{F}(\rho_\omega)=2\sum_{nm} |\langle\psi_m|\partial_{\omega}\rho_{\omega}|\psi_n\rangle|^2/(\lambda_n+\lambda_m)$, where $\lambda_i$ and $|\psi_i\rangle\ (i=n,m)$ represent the eigenvalues and eigenvectors of $\rho_\omega$, the sums include only terms with $\lambda_n+\lambda_m\neq 0$. This time the 2-qubit Bell state can beat the SQL. In addition, the 4-qubit GHZ state can even beat the noiseless SQL bound which is equal to 1.

\begin{figure}[tb]
\centering
\includegraphics[width=0.97\textwidth]{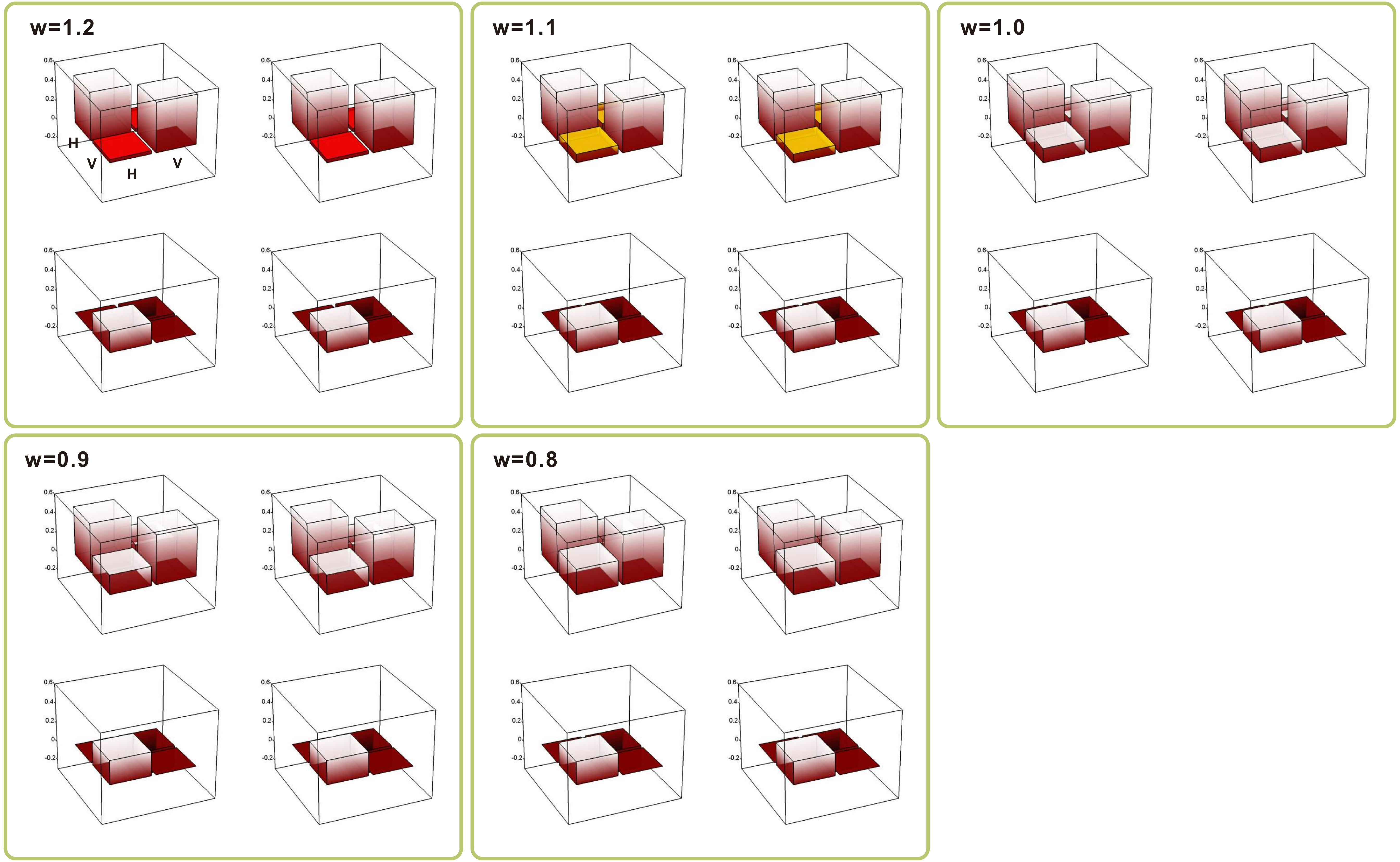}
\caption{\label{Fig:S7} Tomographic results for N=1. In each box, the left two pictures show the real (top) and imaginary (bottom) part of the experimental reconstructed density matrix, the right pictures show the theoretical density matrix. The fidelity of the states are $\{0.9997, 0.9999, 0.9997, 1.0000, 0.9999\}$ respectively.}  \end{figure}

\begin{figure}[tb]
\centering
\includegraphics[width=0.97\textwidth]{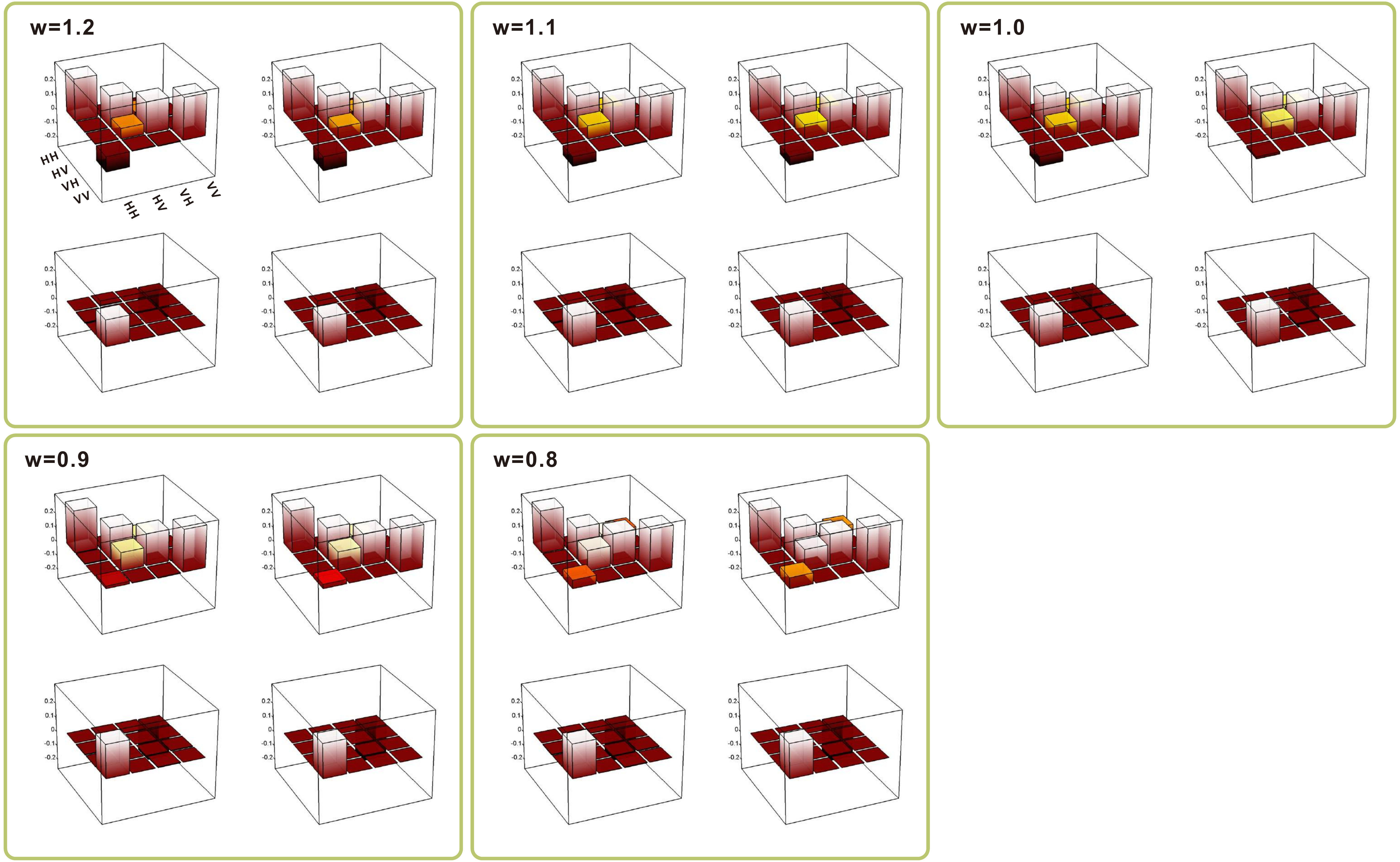}
\caption{\label{Fig:S8}Tomographic results for N=2, the fidelity of the states are $\{0.9991, 0.9993, 0.9996, 0.9992, 0.9989\}$ respectively.}  \end{figure}

\begin{figure}[tb]
\centering
\includegraphics[width=0.97\textwidth]{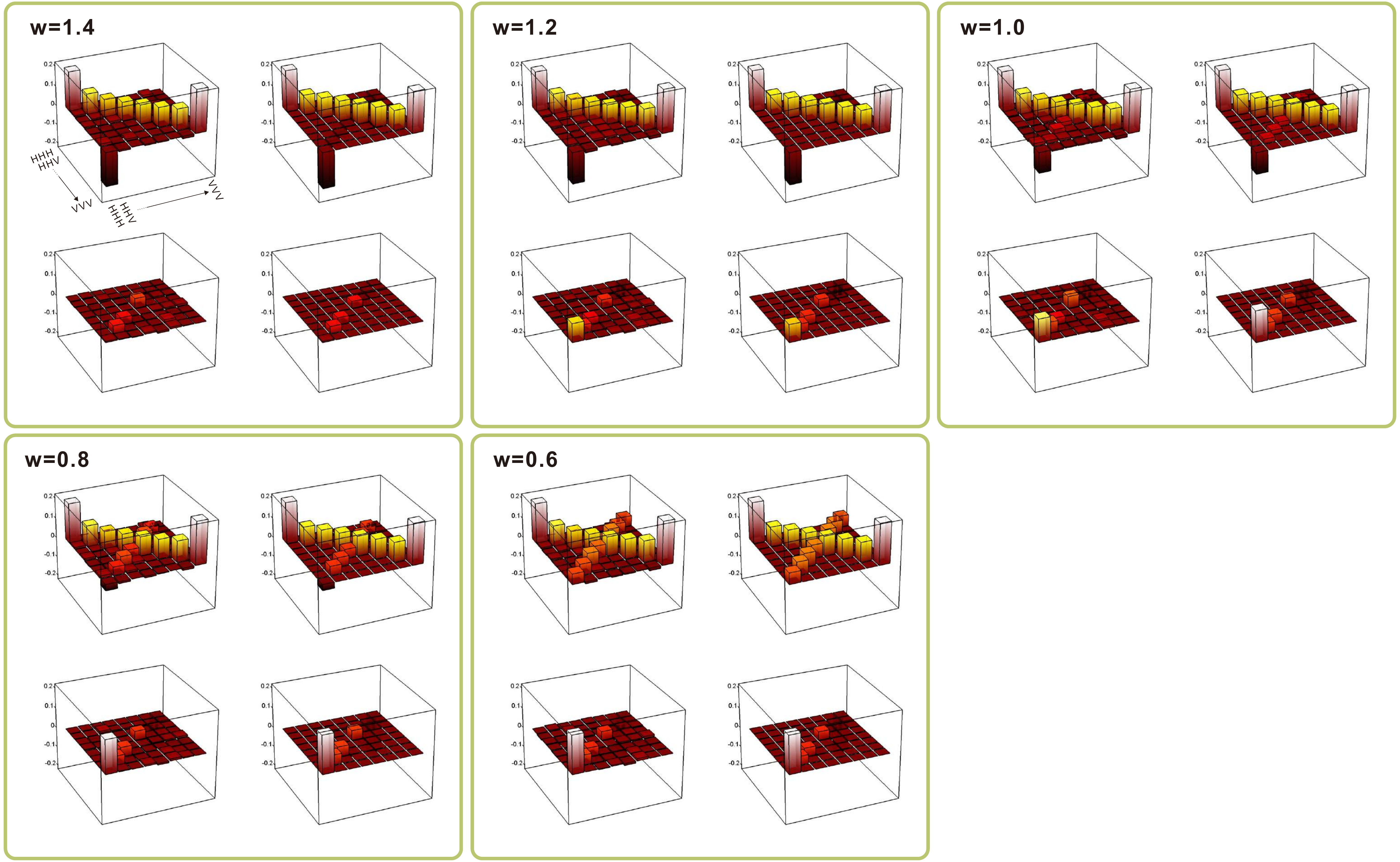}
\caption{\label{Fig:S9}Tomographic results for N=3, the fidelity of the states are $\{0.9879, 0.9884, 0.9779, 0.9815, 0.9804\}$ respectively.}  \end{figure}

\begin{figure}[tb]
\centering
\includegraphics[width=0.97\textwidth]{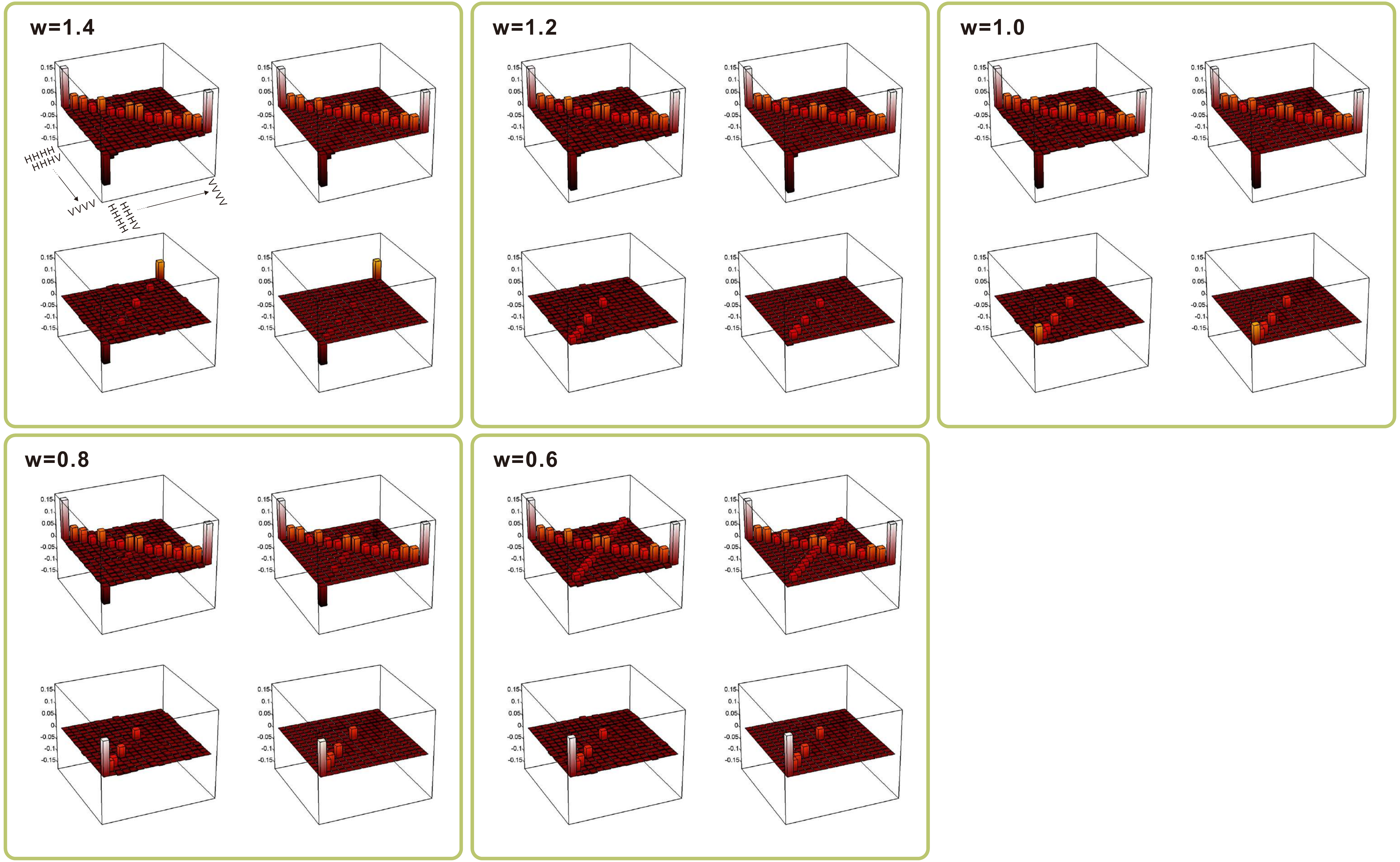}
\caption{\label{Fig:S10}Tomographic results for N=4, the fidelity of the states are $\{0.9466, 0.9586, 0.9644, 0.9643, 0.9612\}$ respectively.}  \end{figure}

\begin{figure}[tb]
\centering
\includegraphics[width=0.75\textwidth]{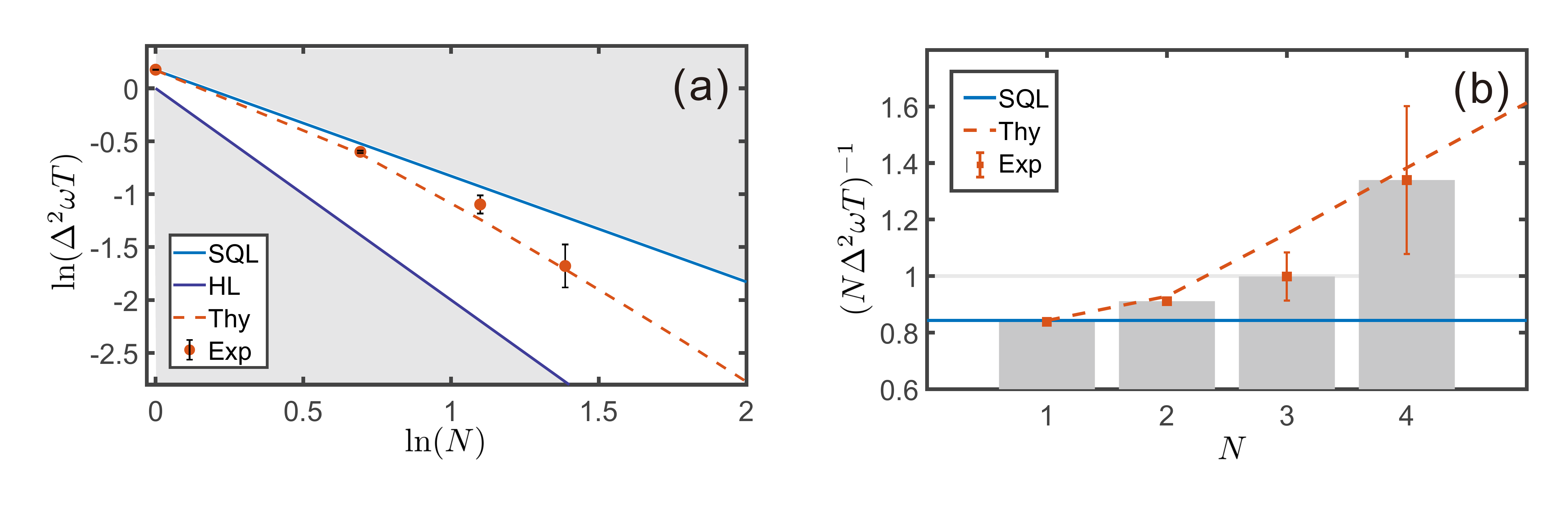}
\caption{\label{Fig:S11} (a) The log-log plot of the precision scaling determined by quantum fisher information. (b) Quantum fisher information per photon as a function of N.}  \end{figure}

%
%

\clearpage
\end{widetext}

\end{document}